\documentclass[twocolumn,longbibliography]{revtex4-2}
\usepackage{amsmath,amsfonts,amssymb,amsthm,epsfig,array,graphicx,subfigure}
\usepackage{titlesec}
\usepackage{feynmp}
\usepackage{setspace}
\usepackage{comment}
\usepackage{dsfont}
\usepackage{color}
\usepackage[pdftex,colorlinks=true,linkcolor=blue,citecolor=blue]{hyperref}
\begin{document}
\title{Reentrant localization transition in the Su-Schrieffer-Heeger model with random-dimer disorder}

\author{Zheng-Wei Zuo}
\author{Dawei Kang}
\affiliation{School of Physics and Engineering, and Henan Key Laboratory of Photoelectric Energy Storage Materials and Applications, Henan University of Science and Technology, Luoyang 471023, China}
\date{\today }

\begin{abstract}
The question of whether the topological insulators can host a stable nontrivial phase in the presence of spatially correlated disorder is a fundamental question of interest. Based on theoretical investigations, we analyze the effect of random-dimer disorder on the quantum phase transitions of the Su-Schrieffer-Heeger model. We explicitly demonstrate that, due to the absence of symmetry, there are no in-gap edge states in certain disordered topological nontrivial gapped phase and the bulk-boundary correspondence breaks down. However, the fractionalized end charges still appear at the ends of the chain. The energy distribution possibility of the in-gap edge states and the possible values of fractionalized end charges are dependent on the concentration of random-dimer disorder. On the other hand, random-dimer disorder and dimer hopping intertwine in an interesting manner. Reentrant localization transition behavior appears, which is evidenced by the fingerprint of the inverse participation ratio, normalized participation ratio, and tunneling conductivity.

\end{abstract}

\maketitle

\section{Introduction}
Research on topological quantum states has been one of the fascinating and influential fields in condensed-matter physics\cite{Bernevig13Book, MoessnerR21Book, Hasan10RMP, QiXL11RMP, Franz15RMP, WenXG17RMP, BergholtzEJ21RMP}. As is well known, the discoveries of the integer and fractional quantum Hall effects\cite{Klitzing80PRL, Tsui82PRL} have triggered numerous theoretical explanations and experimental explorations of topological quantum matters. In particular, findings on topological insulators, topological superconductors, and topological semi-metals comprise the major themes at present. More recently, extending the notions of topological quantum states to the high-order topological phases\cite{Benalcazar17SCI,Benalcazar17PRB,Langbehn17PRL,SongZD17PRL,SchindlerF18SR}, topological crystalline phases\cite{FuL11PRL,Slager12NTP,KruthoffJ17PRX,PoHC17NTC,BradlynB17NT}, and non-Hermitian topological phases\cite{BergholtzEJ21RMP} has become the broad frontier of current research. Usually, topological quantum matter exhibits nontrivial edge or surface states, which have potential applications from spintronics to topological quantum computation\cite{Nayak08RMP}.

Disorders are almost unavoidable in real quantum materials, which dramatically affects physical properties such as conductivity due to impurity scattering. Thus, the Anderson transitions between localized and metallic phases in the disordered systems have been intensively studied\cite{Anderson58PR, AbrahamsE79PRL, FerdinandE08RMP}. For the dimensions $D \leq 2$, the scaling theory of localization\cite{AbrahamsE79PRL} predicts that all single-particle states are localized for non-interacting fermions with uncorrelated disorder, regardless of the amount of disorder. For spatially correlated disorders, exception situations can occur. For example, the random-dimer model\cite{DunlapDH90PRL, PhillipsP91SCI, WuHL91PRL}, in which the onsite potentials for pairs of lattice sites are assigned one of two values at random, exhibits surprising localization-delocalization transitions, which has also been demonstrated experimentally\cite{BellaniV99PRL, NaetherN13NJP}. On the other hand, the effect of correlated disorder on topological insulators\cite{MondragonShem14PRL, AltlandA14PRL2, GirschikA13PRB, WautersMM19PRL, HaywardALC21PRA} has been vividly investigated recently. Strong disorder can transform such systems into topological Anderson insulators\cite{LiJ09PRL, GrothCW09PRL, GuoHM10PRL, MeierEJ18SCI,ZhangDW20SC,LuoXW19arXiv}. As a one-dimensional topological insulator, the Su-Schrieffer-Heeger (SSH) model\cite{SuWP79PRL} provides a paradigmatic example of the most spectacular topological phases. There are two nontrivial in-gap edge states and a fractionalized end charge $e/2$ at half filling in the topological phase of the finite SSH model. The effect of diagonal and off-diagonal disorder on nontrivial quantum phase transitions of the SSH model has also been analyzed\cite{MeierEJ18SCI, LiuT18PLA, LonghiS20OL, BeatrizP19PRB, ScollonM20PRB, ShilpiR21PRL}.

It is interesting to investigate the effect of spatially correlated disorder such as random-dimer disorder on the SSH model, which has not been explored and can be easily realized in an optical waveguide array. Our numerical calculations demonstrate that the interplay of dimer hopping and correlated disorder can induce various unexpected features. Our results reveal that (i) correlated disorder can cause the topological phase without the in-gap edge states; (ii) the energy distribution possibility of the in-gap edge states and possible values of the fractionalized end charge depend on the concentration of random-dimer disorder; (iii) the quantum transition is not tied to a delocalization-localization transition of the eigenstates; (iv) the reentrant localization transition behaviors appear in the non-topological parameter region; and (v) the topological phase transition regions depend on the concentration of random-dimer disorder, but the localization-delocalization transition does not. The rest of this paper is organized as follows. First, in Sec.\ref{Model}, we present the system model. Then, in Sec. \ref{HalfFilling}, the quantum phase transitions at half filling are discussed. In Sec. \ref{LDtransition}, the localization-delocalization transition is investigated. The conclusion is given in Sec. \ref{Conclusion}.

\section{Model and Result}
\label{Model}

We consider a one-dimensional model of spinless fermions with nearest-neighbor dimer hoppings and random-dimer disorder on an $L$ sites chain. The tight-binding Hamiltonian of the system is 
\begin{equation}
H=\sum_{j=1}^{L-1}t\left[1-(-1)^{j} \Delta\right] (c_{j}^{\dagger} c_{j+1}+\mathrm{H.c.}) +\sum_{j=1}^{L} \epsilon_{i}n_{j}
\end{equation}
where $t$ is the hopping amplitude (set to unit of energy hereafter), and $\Delta$ denotes the dimerization parameter. $c_{j}^{\dagger} (c_{j}) $ creates (annihilates) a fermionic particle at site $j$. $n_{j}=c_{j}^{\dagger} c_{j}$ is the electronic occupation number operator. For the onsite potential, $\epsilon_{b}$ ($\epsilon_{a}$) is assigned at random to pairs of lattice sites (that is, two sites in succession) with probabilities $q$ and $1-q$, respectively. The model proposed herein can be easily realized in an optical waveguide array\cite{NaetherN13NJP}. The one-dimensional waveguide array contains two types of waveguides with propagation constants $\epsilon_{a}$ and $\epsilon_{b}$, which are assigned at random to pairs of dimer lattice sites (two adjacent waveguides with the same onsite potential). The probability of onsite potential $\epsilon_{b}$ ($\epsilon_{a}$) is $q$ ($1-q$). The two staggered values of the inter-waveguide separation are set, so that the coupling strengths $t(1-\Delta)$ and $t(1+\Delta)$ can be easily achieved. For $\epsilon_{i}=0$, the system reduces to the SSH model\cite{SuWP79PRL}. The system becomes the conventional random-dimer model\cite{DunlapDH90PRL,PhillipsP91SCI} when the dimerization $\Delta=0$. For simplicity, in the following we set $q=0.5$ and $\epsilon_{a}=0$.

\subsection{Phase transitions at half filling}\label{HalfFilling}

\begin{figure}[tpbh]
\includegraphics[scale=0.92]{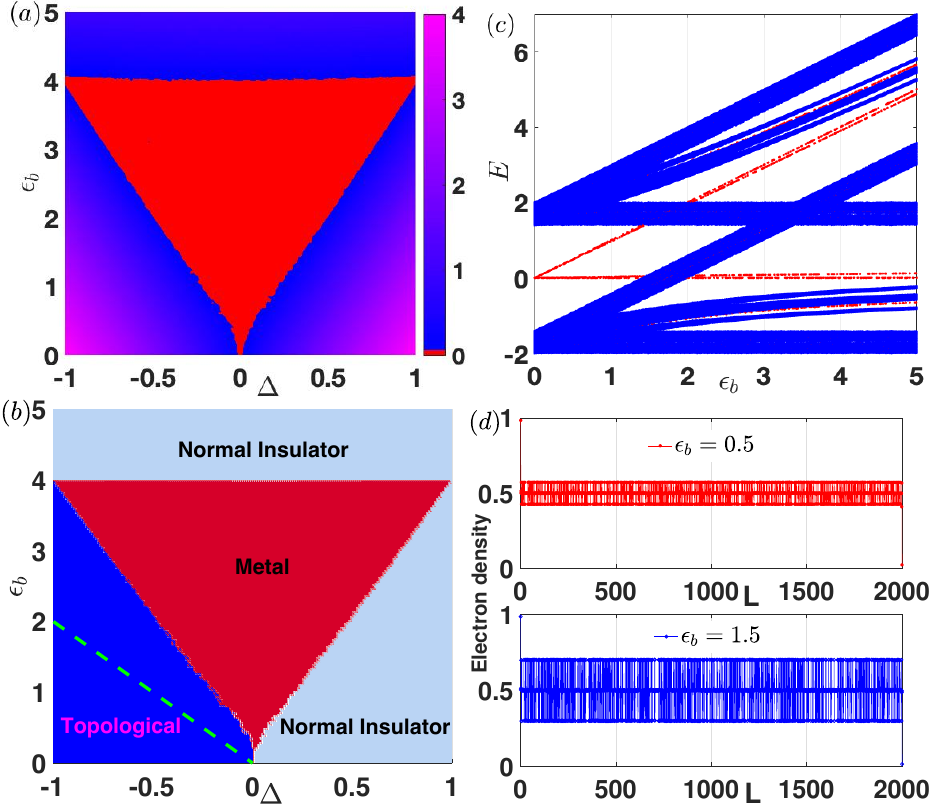}
\caption{(a) Energy gap of the system as a function of dimerization parameter ($\Delta$) and onsite potential ($\epsilon_{b}$) under periodic boundary condition. (b) Phase diagram according to polarization calculated in real space. (c) Energy spectrum as a function of onsite potential $\epsilon_{b}$ with dimerization amplitude $\Delta=-0.7$. Blue regions are the energy spectrum with periodic boundary condition. Red dots are edge states under open boundary condition. (d) Electron density at $\epsilon_{b}=0.5$ and $1.5$ with $\Delta=-0.7$. Other parameters are $L=2000$ at half filling under each disorder configuration.}
\label{FigHalffilling}
\end{figure}

As is well known, the SSH model at half filling is in the topological (non-topological) phase when the dimerization strength $\Delta<0 $ ($>0$)\cite{ShenSQ12book,AsbothJ16}. The appearance of random-dimer disorder breaks the translation invariance. Thus, we investigate the topological properties of the system in real space. First, the energy gap under periodic boundary condition as a function of dimerization parameter and random-dimer disorder is shown in Fig.\ref{FigHalffilling}a when the system size is $L=2000$ at half filling under one realization. In the center red region, the system is gapless because of bulk-gap closing. In other regions, the system is in the insulating phase. Here, we use Resta’s formula of polarization\cite{Resta98PRL} in real space as the topological invariant with which to characterize the topological phase transitions in this disordered system, which is defined as
\begin{equation}
P=\frac{1}{2 \pi} \operatorname{Im}\left[\ln \left\langle \Psi\left| e^{2 \pi i \sum_{j} \frac{n_{j}}{L}}\right| \Psi\right\rangle\right]  \quad\text{(mod 1)}
\end{equation}
where the operator $n_{j}$ is electron number operator at site $j$, and $\Psi$ is the many-body ground state.
 
According to the numerical calculations of the polarization, we can obtain the phase diagram at half filling, as shown in Fig.\ref{FigHalffilling}b.  For the lower left (blue) region, the polarization $P=0.5$, which indicates that the system is in the nontrivial topological phase. In the center (red) region, the system is in the metal phase. In the lower right region, the system is in the trivial insulating phase with polarization $P=0$. The polarization $P$ does not take a fixed value for the upper region ($\epsilon_{b}>4$) under more than 200 realizations. Compared with Fig\ref{FigHalffilling}a, it can be seen that the topological phase transitions coincide with the energy-gap closing. The topological states are robust against weak disorder, and the more strongly the dimer parameter $\Delta$ evolves, the larger the threshold of disorder strength $\epsilon_{b}$ at the phase transition becomes. 
 
 Next, we investigate the fate of the in-gap edge states in the topological phases. In Fig.\ref{FigHalffilling}c, the energy spectra as functions of disorder strength $\epsilon_{b}$ with the dimerization amplitude $\Delta=-0.7$ are plotted. As shown in Ref.\cite{RhimJW18PRB}, the edge states in one-dimensional band insulators have inherent fragility because they are separated from bulk bands, except those protected by symmetry. In the system studied herein, the chiral and inversion symmetries are lost because of the spatially random dimer-disorder and the zero-energy in-gap states of the clean system become energetic. From Fig.\ref{FigHalffilling}c, it can be seen that for small disorder the two in-gap edge states survive in the energy gap at half filling, and the degeneracy is lifted due to random-dimer disorder. The two in-gap edge states begin to assimilate into the bulk states when the disorder strength $\epsilon_{b}$ passes a critical value (approximately $1.4$). In other words, there are no in-gap edge states in the nontrivial topological gapped phase and the bulk–boundary correspondence breaks down. The correlated disorder provides another route to the breakdown of bulk–boundary correspondence. According to the presence or absence of in-gap edge states, the green dashed curve in Fig.\ref{FigHalffilling}b divides the topological phase region into two parts. For the upper region, there are no in-gap edge states in this topological phase. As the disorder strength $\epsilon_{b}$ increases ($1.4<\epsilon_{b}<4$), the energy gap closes and the system remains in the metal phase, as shown in Fig.\ref{FigHalffilling}c. When the on-site potential $\epsilon_{b}$ goes beyond $4$, the system enters the trivial insulating phase without in-gap edge states at half filling.

 \begin{figure}[t]
\includegraphics[scale=0.34]{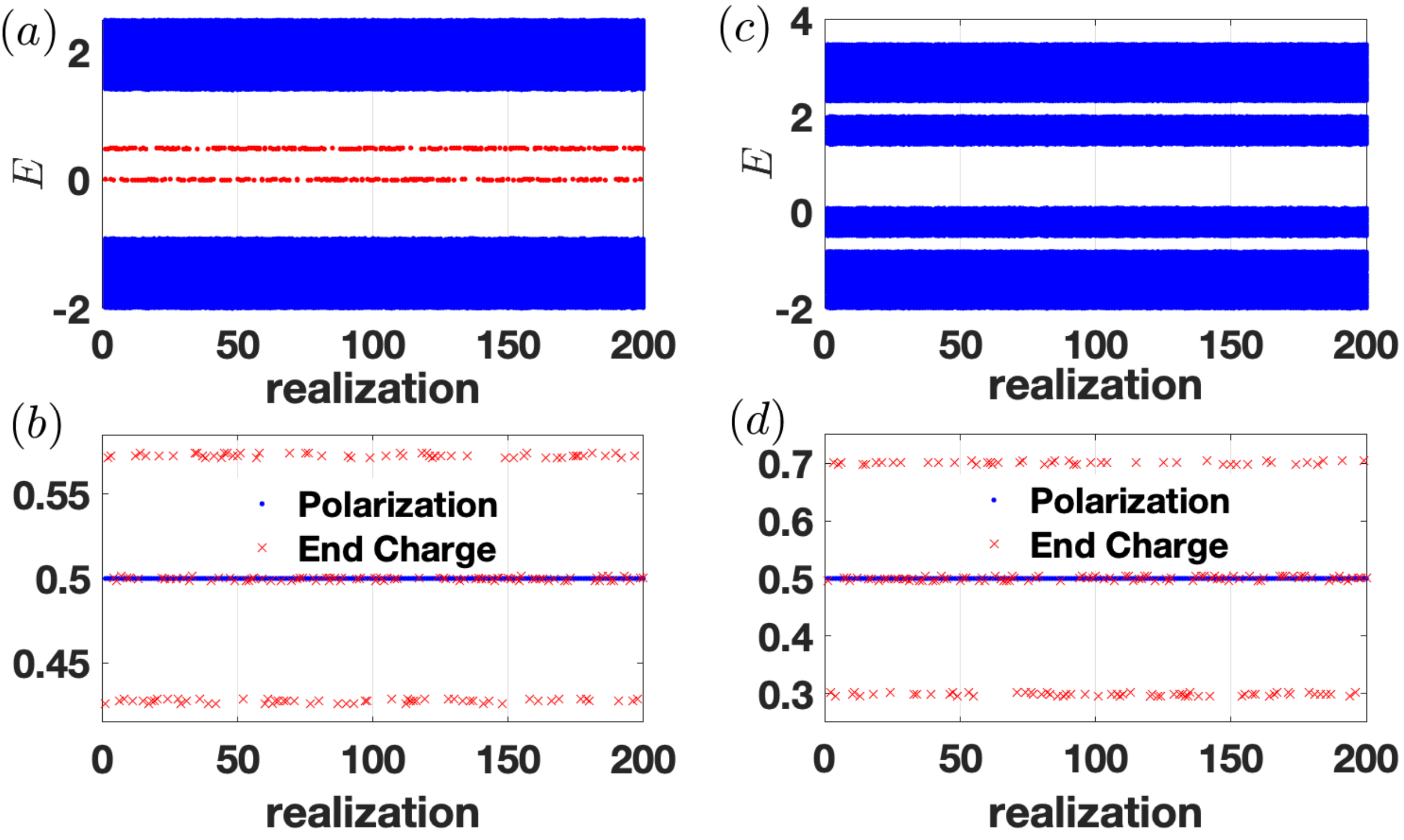}
\caption{Energy spectra, polarization and left fractionalized end charges at half filling under 200 realizations at $\epsilon_{b}=0.5$ [(a) and (b)] and $1.5$ [(c) and (d)]. Other parameters are $L=2000$ and $\Delta=-0.7$.}
\label{FigEndCharge}
\end{figure}
 
 The fractionalized charges appear at the ends of the chain as a hallmark of the topological states in the one-dimensional chain. It is interesting to analyze the end-charge properties of the SSH model with correlated disorder. The electron density with $\epsilon_{b}=0.5$ and $1.5$ at half filling under the open boundary condition is shown in Fig.\ref{FigHalffilling}d. According to the  formula $Q_L=\sum_{j=1}^{L/2}(\langle c_{j}^{\dagger}c_{j}\rangle-\bar{\rho})$ ($\bar{\rho}$ is the bulk charge density) of the left end charge, we numerically found that the end charges in the topological phase are fractional at a nonzero disorder strength of $\epsilon_{b}$. In Fig.\ref{FigEndCharge}, we plot the energy spectrum, polarization, and left fractionalized end charge at half filling under 200 realizations at $\epsilon_{b}=0.5$ and $1.5$. For the in-gap edge states of topological states (red dots in Fig.\ref{FigEndCharge}a), numerical simulations indicate that the possibility of the high-energy (low-energy) in-gap edge state approaches the possibility $q$ ($1-q$) of $\epsilon_{b}$ ($\epsilon_{a}$) under extreme realizations. In contrast, the possibility of the fractionalized end charge $e/2$ tends towards the possibility $(1-q)$ of $\epsilon_{a}$, while the total possibility of the small and large fractionalized end charge (not equal to $e/2$) approaches the possibility $q$ of $\epsilon_{b}$ (see Fig.\ref{FigEndCharge}c for detail). This behavior of the fractionalized end charge also applies to the case of the topological phase without in-gap edge states (Fig.\ref{FigEndCharge}d). The possibility properties of in-gap edge states and fractionalized end charge are unique to the topological phase of the system studied here. The average value of the left fractionalized end charge under 200 realizations approaches the stable value $e/2$. In addition, it should be stressed that the other edge states, due to the on-site disorder, appear when the filling fraction deviates from the half filling case. However, the numerical calculations of polarization do not take the fixed value under distinct realizations and the average value exhibits uncertainty.
 
 The probability-dependent energy distribution of the in-gap edge states can be understood by the following argument: When the system is in topological nontrivial states at half filling, the energy location of an in-gap (assumed to be locating at the left edge) state depends on the dimer disorder $\epsilon_{a}$ or $\epsilon_{b}$ sited at the two left end sites under a disorder configuration. When the dimer disorder $\epsilon_{b}$ is located at the two left end sites, the in-gap left edge state occupies the high-energy region of the energy gap. Thus, the energy distribution possibility of the high-energy (low-energy) in-gap edge state depends on the concentration of random-dimer disorder $\epsilon_{b}$ ($\epsilon_{a}$). As the onsite potential $\epsilon_{b}$ increases further (larger then the critical value), the in-gap edge states assimilate into the bulk states. As long as the energy gap does not close, the system still remains in the topological phase, while there are no in-gap edge states.

In short, the numerical simulations demonstrate the existence of the topological insulating phase without the in-gap edge states at half filling induced by random-dimer disorder. The energy distribution possibility of the in-gap edge states and possible values of the fractionalized end charge depend on the concentration of random-dimer disorder.

\subsection{Localization-delocalization transition}\label{LDtransition}

We next investigate the localization-delocalization transition of interest due to the spatially correlated disorder. A way of discerning between localized and extended states is provided by investigating the inverse participation ratio (IPR) and normalized participation ratio (NPR), which are given by
\begin{equation}
\mathrm{IPR_n}=\sum_{j=1}^{L} \left|\phi_j^n\right|^4, \mathrm{NPR_n}=(L\sum_{j=1}^{L} \left|\phi_j^n\right|^4)^{-1}
\end{equation}
respectively, where $\phi_j^n$ is the $n$th eigenstate. The IPR of an extended state scales as $1/L$, thereby vanishing in the large-$N$ limit, while remaining finite for a localized state. For the intermediate states, the mean $\langle\mathrm{IPR}\rangle$ (the average of IPR for all the eigenstates) and $\langle\mathrm{NPR}\rangle$ are finite. The localization-delocalization phase diagram can be obtained by computing the $\eta$ quantity\cite{LiX20PRB}: 
\begin{equation}
\eta=\log _{10}[\langle\mathrm{IPR}\rangle \times\langle\mathrm{NPR}\rangle]
\end{equation}

\begin{figure}[b]
\includegraphics[scale=0.92]{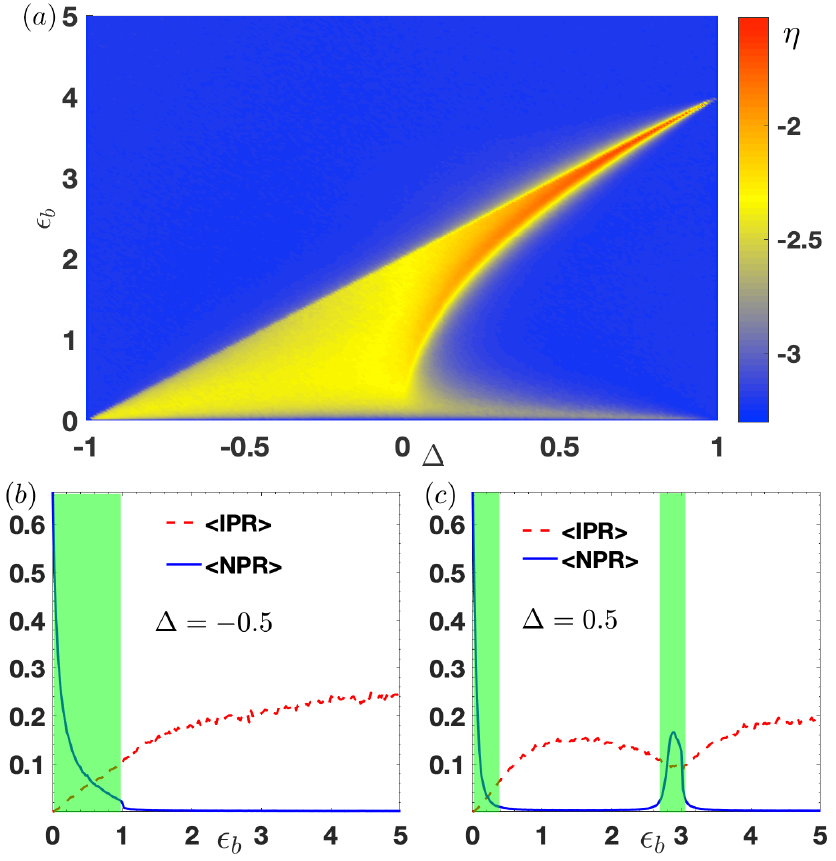}
\caption{(a) $\eta$ quantity of the system as a function of dimerization parameter ($\Delta$) and onsite potential ($\epsilon_{b}$) under periodic boundary condition. Mean $\langle\mathrm{IPR}\rangle$ and $\langle\mathrm{NPR}\rangle$ as a function of $\epsilon_{b}$ with $\Delta=-0.5$ (b) and 0.5 (c). System size in simulation is $L=2000$.}
\label{FigPhasediagram}
\end{figure}

From Fig.\ref{FigPhasediagram}a, it can be seen that away from the clean limit the localization transition occurs through a small disorder and the system hosts the mobility edge in both the topological trivial and nontrivial regions. We first investigate the topological regime($\Delta<0$). Although the occupied states of the system are topological at half filling, the random-dimer order can change the location properties of the occupied states as shown in Fig. \ref{FigPhasediagram}b. The finite $\langle\mathrm{IPR}\rangle$ and $\langle\mathrm{NPR}\rangle$ signify the presence of both the localized and extended states(see the green shaded region in Fig.\ref{FigPhasediagram}b), and the system enters into the coexisting phase. When the on-site potential $\epsilon_{b}$ goes beyond the critical values, all the states of the system will be localized. For example, $\langle\mathrm{IPR}\rangle\neq 0$ and $\langle\mathrm{NPR}\rangle=0$ for all the onsite potentials $\epsilon_{b}>1$ when the dimer strength $\Delta=-0.5$(see Fig\ref{FigPhasediagram}b). Moreover, as the dimer strength $|\Delta|$ increases, the critical value of the on-site potential $\epsilon_{b}$ decreases. In addition, compared with Figs.\ref{FigPhasediagram}a and \ref{FigHalffilling}b, it can be concluded that topological quantum transition is not tied to a localization-delocalization transition of the eigenstates.

\begin{figure}[h]
\includegraphics[scale=0.34]{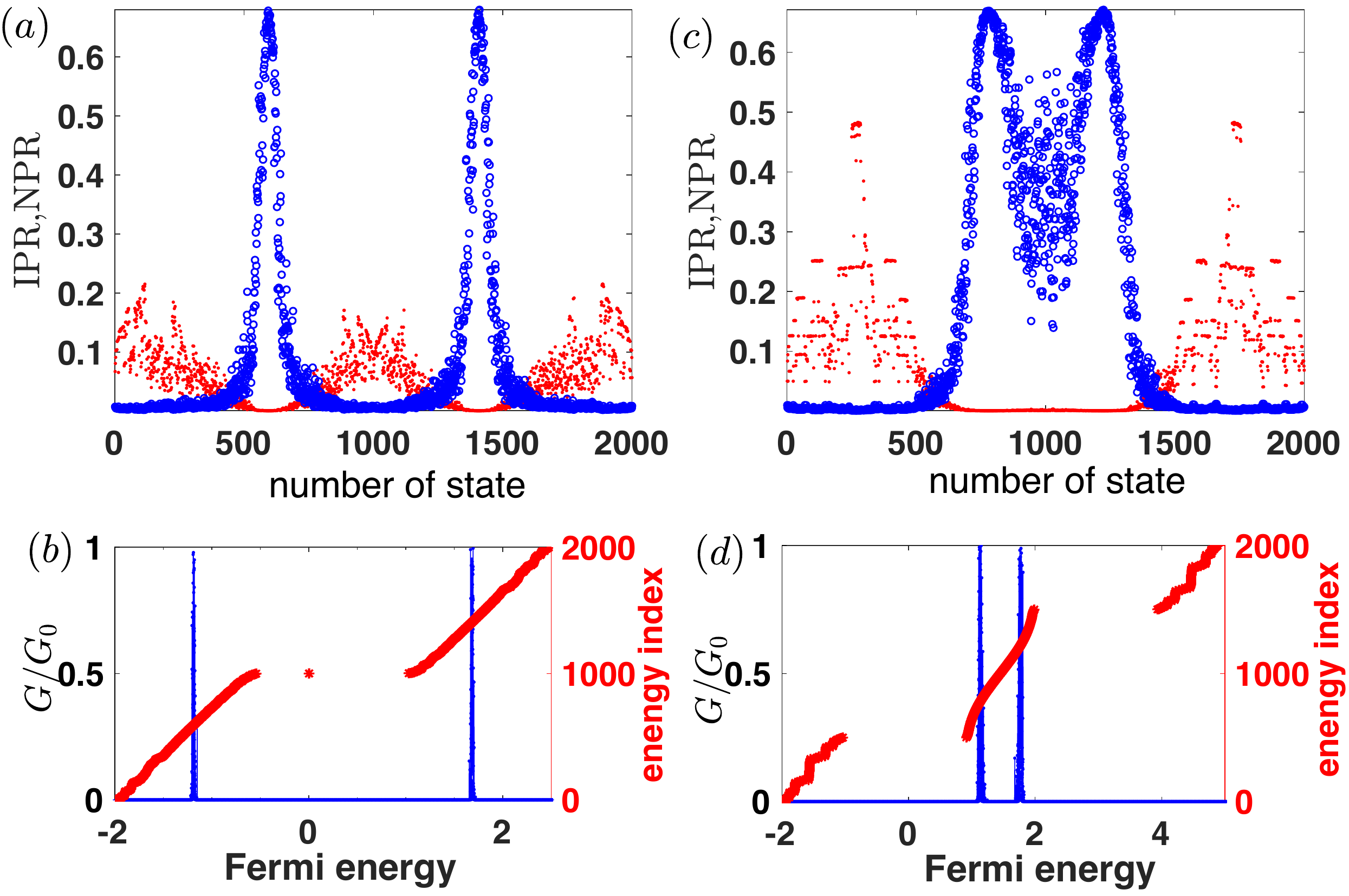}
\caption{(a) and (c) The IPR (red dots), and NPR (blue open circles) of all eigenstates with system size $L=2000$ under open boundary condition. (b) and (d) Tunneling conductance (blue curve) and energy spectrum (red asterisks) as a function of Fermi energy. Other parameters are $\Delta=-0.5$ and $\epsilon_{b}=0.5$ [(a) and (b)] and $\Delta=0.5$, and $\epsilon_{b}=2.9$ [(c) and (d)] .}
\label{FigTransport}
\end{figure}

In the non-topological regime($\Delta>0$), the system first undergoes the localization-delocalization transition due to the very small disorder, similar to the behavior in the topological regions. All the eigenstates of the system are localized when the onsite potential $\epsilon_{b}$ goes beyond small critical values, as shown in Figs.\ref{FigPhasediagram}a and \ref{FigPhasediagram}c. Surprisingly, as the onsite potential $\epsilon_{b}$ increases, the system again undergoes the localization-delocalization transition, which is a counter-intuitive result. The second green region (at approximately $\epsilon_{b}=3$) of Fig.\ref{FigPhasediagram}c indicates the reentrant localization phenomenon, in which the $\langle\mathrm{IPR}\rangle$ and $\langle\mathrm{NPR}\rangle$ are nonzero. Next, we use the IPR, NPR, and transport properties to analyze the coexisting phase of the localized and extended states in the reentrant regime. In Fig\ref{FigTransport}c, the $\mathrm{IPR_n}$ and $\mathrm{NPR_n}$ for all eigenstates are plotted. Clearly, for some eigenstates, $\mathrm{IPR_n}=0$ and $\mathrm{NPR_n}\neq 0$, with which the extended properties of the eigenstates are encoded. As discussed in the random-dimer model\cite{DunlapDH90PRL}, the numerical simulations demonstrate that $\sqrt L$ of all eigenstates in the coexisting phase are extended and unscattered, which is attributed to the single-dimer impurity resonance. To further investigate the quantum-transport properties of these extended states, we calculated the tunneling conductivity by using the scattering theory\cite{GrothCW14NJP}. To this end we attach the first and last sites of the chain to two different perfect leads. The hoppings between the chain and leads are set to $t=1$. The tunneling conductance is shown in Fig.\ref{FigTransport}d, when the parameters $\Delta=0.5$ and $\epsilon_{b}=2.9$. Obviously, at the Fermi energy of the extended states, the tunneling conductance approaches the quantized conductance $G_0=2e^2/h$. For comparison, the IPR, NPR, and tunneling conductance for the parameters $\Delta=-0.5$ and $\epsilon_{b}=0.5$ in the coexisting region of topological regime are also plotted in Fig.\ref{FigTransport}a and \ref{FigTransport}b. In brief, the IPR, NPR, and tunneling conductance provide compelling evidence for the localization-delocalization transition in the non-topological regions. This reentrant localization transition behavior is due to the interplay of dimer hopping and random-dimer disorder. 

\begin{figure}[h]
\includegraphics[scale=0.9]{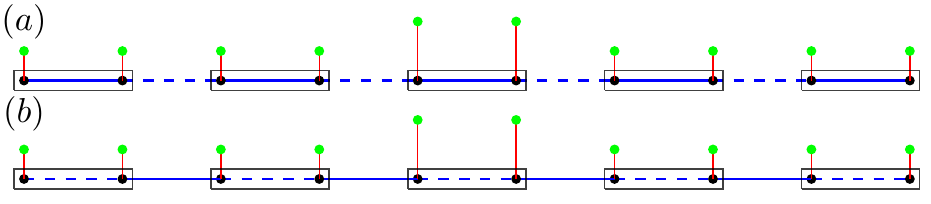}
\caption{Two geometrical configurations for a dimer defect (fixed on lattice sites 5 and 6) in the SSH model with system size $L=10$. The box containing two lattice sites denotes the unit cell. Solid and dotted blue horizontal lines are alternatively dimerized hoppings. Red vertical lines indicate the onsite potentials $\epsilon_{a}$ (short lines) and $\epsilon_{b}$ (long lines).}
\label{Fig5}
\end{figure}

Herein, we use a random-dimer defect $\epsilon_b$ to analyze the two different behaviors of localization-delocalization transitions in topological and non-topological regimes. Because both of the on-site potentials $\epsilon_a$ and $\epsilon_b$ are random in pairs of lattice sites, each disordered dimer defect can only stay on a whole unit cell of the clean SSH model (stretching across the two unit cells is not allowed). The dimer defect $\epsilon_b$ can be placed on lattice sites $2k-1$ and $2k$ ($k$ is a non-negative integer and smaller than $L/2$ ), and all other pairs of lattice sites along the chain are assigned to $\epsilon_a$.  Fig. \ref{Fig5} shows two geometrical configurations for the SSH model with system size $L=10$ as an example, where a dimer defect $\epsilon_b$ is fixed on lattice sites 5 and 6. We can further derive that in thermodynamic limit there are only two types of geometrical configurations for the fixed dimer defect and dimerization hoppings, which depends on the sign of the dimerization hopping parameter $\Delta$. The two configurations induce the different localization properties of the quantum states. Thus, the two different behaviors of localization-delocalization transitions induced by the alternatively dimerized hopping in topological and non-topological regimes can be understood more comprehensively.

A comment is in order at this point. Because the energy gap at half filling can be changed by the concentration of random-dimer disorder, it is easily verified that the regions of phase diagram at half filling discussed in Sec. \ref{HalfFilling} depend on the concentration $p$ of the onsite potential $\epsilon_{b}$. However, the conclusions regarding the localization-delocalization transition in Sec. \ref{LDtransition} are independent of the concentration $p$ and only rely on the relative difference of $\left|\epsilon_{a}-\epsilon_{b}\right|$.

\section{Conclusion}\label{Conclusion}

in this paper we investigated the effect of random-dimer disorder on quantum phase transitions in the disordered SSH model, which leads to phenomena drastically distinct from the clean system. We used the topological invariant, edge states, and fractional end charge to analyze the existence of the topological phase without in-gap edge states at half filling. The energy distribution possibility of the in-gap edge states and possible values of the fractionalized end charge depend on the concentration of random-dimer disorder. For the localization feature of the system, the reentrant localization phenomenon in the topological trivial regime appears and is revealed in detail by the IPR, NPR, and tunneling conductivity. Our findings could open up new directions to study and facilitate understanding of the competition mechanism of the modulated hopping and correlated disorder. Moreover, the non-Hermiticity could also lead to delocalization in one dimension\cite{KawabataK21PRL}. It would be interesting to investigate the quantum phase transitions of the non-Hermitian SSH model with spatially correlated disorder.

\textit{Note added}: Recently, we became aware of a related work about the SSH model with random binary disordered hoppings in Ref.~\cite{LiuSH22PLA}.

\emph{Acknowledgements.}--Z.W.Z. is grateful to X. Deng for helpful discussion. This work was supported by the National Natural Science Foundation of China (Grant Nos. 12074101, and 11604081). Z.W.Z. is also sponsored by the Natural Science Foundation of Henan (Grant No. 212300410040).


\begin{thebibliography}{52}%
\makeatletter
\providecommand \@ifxundefined [1]{%
 \@ifx{#1\undefined}
}%
\providecommand \@ifnum [1]{%
 \ifnum #1\expandafter \@firstoftwo
 \else \expandafter \@secondoftwo
 \fi
}%
\providecommand \@ifx [1]{%
 \ifx #1\expandafter \@firstoftwo
 \else \expandafter \@secondoftwo
 \fi
}%
\providecommand \natexlab [1]{#1}%
\providecommand \enquote  [1]{``#1''}%
\providecommand \bibnamefont  [1]{#1}%
\providecommand \bibfnamefont [1]{#1}%
\providecommand \citenamefont [1]{#1}%
\providecommand \href@noop [0]{\@secondoftwo}%
\providecommand \href [0]{\begingroup \@sanitize@url \@href}%
\providecommand \@href[1]{\@@startlink{#1}\@@href}%
\providecommand \@@href[1]{\endgroup#1\@@endlink}%
\providecommand \@sanitize@url [0]{\catcode `\\12\catcode `\$12\catcode
  `\&12\catcode `\#12\catcode `\^12\catcode `\_12\catcode `\%12\relax}%
\providecommand \@@startlink[1]{}%
\providecommand \@@endlink[0]{}%
\providecommand \url  [0]{\begingroup\@sanitize@url \@url }%
\providecommand \@url [1]{\endgroup\@href {#1}{\urlprefix }}%
\providecommand \urlprefix  [0]{URL }%
\providecommand \Eprint [0]{\href }%
\providecommand \doibase [0]{https://doi.org/}%
\providecommand \selectlanguage [0]{\@gobble}%
\providecommand \bibinfo  [0]{\@secondoftwo}%
\providecommand \bibfield  [0]{\@secondoftwo}%
\providecommand \translation [1]{[#1]}%
\providecommand \BibitemOpen [0]{}%
\providecommand \bibitemStop [0]{}%
\providecommand \bibitemNoStop [0]{.\EOS\space}%
\providecommand \EOS [0]{\spacefactor3000\relax}%
\providecommand \BibitemShut  [1]{\csname bibitem#1\endcsname}%
\let\auto@bib@innerbib\@empty
\bibitem [{\citenamefont {Bernevig}\ and\ \citenamefont
  {Hughes}(2013)}]{Bernevig13Book}%
  \BibitemOpen
  \bibfield  {author} {\bibinfo {author} {\bibfnamefont {B.~A.}\ \bibnamefont
  {Bernevig}}\ and\ \bibinfo {author} {\bibfnamefont {T.~L.}\ \bibnamefont
  {Hughes}},\ }\href {http://press.princeton.edu/titles/10039.html} {\emph
  {\bibinfo {title} {Topological insulators and Topological Superconductors}}}\
  (\bibinfo  {publisher} {Princeton University Press},\ \bibinfo {year}
  {2013})\BibitemShut {NoStop}%
\bibitem [{\citenamefont {Moessner}\ and\ \citenamefont
  {Moore}(2021)}]{MoessnerR21Book}%
  \BibitemOpen
  \bibfield  {author} {\bibinfo {author} {\bibfnamefont {R.}~\bibnamefont
  {Moessner}}\ and\ \bibinfo {author} {\bibfnamefont {J.~E.}\ \bibnamefont
  {Moore}},\ }\href {https://doi.org/10.1017/9781316226308} {\emph {\bibinfo
  {title} {Topological Phases of Matter}}}\ (\bibinfo  {publisher} {Cambridge
  University Press},\ \bibinfo {year} {2021})\BibitemShut {NoStop}%
\bibitem [{\citenamefont {Hasan}\ and\ \citenamefont
  {Kane}(2010)}]{Hasan10RMP}%
  \BibitemOpen
  \bibfield  {author} {\bibinfo {author} {\bibfnamefont {M.~Z.}\ \bibnamefont
  {Hasan}}\ and\ \bibinfo {author} {\bibfnamefont {C.~L.}\ \bibnamefont
  {Kane}},\ }\bibfield  {title} {\bibinfo {title} {\textit{Colloquium} :
  Topological insulators},\ }\href {https://doi.org/10.1103/RevModPhys.82.3045}
  {\bibfield  {journal} {\bibinfo  {journal} {Rev. Mod. Phys.}\ }\textbf
  {\bibinfo {volume} {82}},\ \bibinfo {pages} {3045} (\bibinfo {year}
  {2010})}\BibitemShut {NoStop}%
\bibitem [{\citenamefont {Qi}\ and\ \citenamefont {Zhang}(2011)}]{QiXL11RMP}%
  \BibitemOpen
  \bibfield  {author} {\bibinfo {author} {\bibfnamefont {X.-L.}\ \bibnamefont
  {Qi}}\ and\ \bibinfo {author} {\bibfnamefont {S.-C.}\ \bibnamefont {Zhang}},\
  }\bibfield  {title} {\bibinfo {title} {Topological insulators and
  superconductors},\ }\href {https://doi.org/10.1103/RevModPhys.83.1057}
  {\bibfield  {journal} {\bibinfo  {journal} {Rev. Mod. Phys.}\ }\textbf
  {\bibinfo {volume} {83}},\ \bibinfo {pages} {1057} (\bibinfo {year}
  {2011})}\BibitemShut {NoStop}%
\bibitem [{\citenamefont {Elliott}\ and\ \citenamefont
  {Franz}(2015)}]{Franz15RMP}%
  \BibitemOpen
  \bibfield  {author} {\bibinfo {author} {\bibfnamefont {S.~R.}\ \bibnamefont
  {Elliott}}\ and\ \bibinfo {author} {\bibfnamefont {M.}~\bibnamefont
  {Franz}},\ }\bibfield  {title} {\bibinfo {title} {\textit{Colloquium} :
  Majorana fermions in nuclear, particle, and solid-state physics},\ }\href
  {https://doi.org/10.1103/RevModPhys.87.137} {\bibfield  {journal} {\bibinfo
  {journal} {Rev. Mod. Phys.}\ }\textbf {\bibinfo {volume} {87}},\ \bibinfo
  {pages} {137} (\bibinfo {year} {2015})}\BibitemShut {NoStop}%
\bibitem [{\citenamefont {Wen}(2017)}]{WenXG17RMP}%
  \BibitemOpen
  \bibfield  {author} {\bibinfo {author} {\bibfnamefont {X.-G.}\ \bibnamefont
  {Wen}},\ }\bibfield  {title} {\bibinfo {title} {Colloquium: Zoo of
  quantum-topological phases of matter},\ }\href
  {https://doi.org/10.1103/RevModPhys.89.041004} {\bibfield  {journal}
  {\bibinfo  {journal} {Rev. Mod. Phys.}\ }\textbf {\bibinfo {volume} {89}},\
  \bibinfo {pages} {041004} (\bibinfo {year} {2017})}\BibitemShut {NoStop}%
\bibitem [{\citenamefont {Bergholtz}\ \emph {et~al.}(2021)\citenamefont
  {Bergholtz}, \citenamefont {Budich},\ and\ \citenamefont
  {Kunst}}]{BergholtzEJ21RMP}%
  \BibitemOpen
  \bibfield  {author} {\bibinfo {author} {\bibfnamefont {E.~J.}\ \bibnamefont
  {Bergholtz}}, \bibinfo {author} {\bibfnamefont {J.~C.}\ \bibnamefont
  {Budich}},\ and\ \bibinfo {author} {\bibfnamefont {F.~K.}\ \bibnamefont
  {Kunst}},\ }\bibfield  {title} {\bibinfo {title} {Exceptional topology of
  non-Hermitian systems},\ }\href
  {https://doi.org/10.1103/RevModPhys.93.015005} {\bibfield  {journal}
  {\bibinfo  {journal} {Rev. Mod. Phys.}\ }\textbf {\bibinfo {volume} {93}},\
  \bibinfo {pages} {015005} (\bibinfo {year} {2021})}\BibitemShut {NoStop}%
\bibitem [{\citenamefont {Klitzing}\ \emph {et~al.}(1980)\citenamefont
  {Klitzing}, \citenamefont {Dorda},\ and\ \citenamefont
  {Pepper}}]{Klitzing80PRL}%
  \BibitemOpen
  \bibfield  {author} {\bibinfo {author} {\bibfnamefont {K.~v.}\ \bibnamefont
  {Klitzing}}, \bibinfo {author} {\bibfnamefont {G.}~\bibnamefont {Dorda}},\
  and\ \bibinfo {author} {\bibfnamefont {M.}~\bibnamefont {Pepper}},\
  }\bibfield  {title} {\bibinfo {title} {New method for high-accuracy
  determination of the fine-structure constant based on quantized Hall
  resistance},\ }\href {https://doi.org/10.1103/PhysRevLett.45.494} {\bibfield
  {journal} {\bibinfo  {journal} {Phys. Rev. Lett.}\ }\textbf {\bibinfo
  {volume} {45}},\ \bibinfo {pages} {494} (\bibinfo {year} {1980})}\BibitemShut
  {NoStop}%
\bibitem [{\citenamefont {Tsui}\ \emph {et~al.}(1982)\citenamefont {Tsui},
  \citenamefont {Stormer},\ and\ \citenamefont {Gossard}}]{Tsui82PRL}%
  \BibitemOpen
  \bibfield  {author} {\bibinfo {author} {\bibfnamefont {D.~C.}\ \bibnamefont
  {Tsui}}, \bibinfo {author} {\bibfnamefont {H.~L.}\ \bibnamefont {Stormer}},\
  and\ \bibinfo {author} {\bibfnamefont {A.~C.}\ \bibnamefont {Gossard}},\
  }\bibfield  {title} {\bibinfo {title} {Two-dimensional magnetotransport in
  the extreme quantum limit},\ }\href
  {https://doi.org/10.1103/PhysRevLett.48.1559} {\bibfield  {journal} {\bibinfo
   {journal} {Phys. Rev. Lett.}\ }\textbf {\bibinfo {volume} {48}},\ \bibinfo
  {pages} {1559} (\bibinfo {year} {1982})}\BibitemShut {NoStop}%
\bibitem [{\citenamefont {Benalcazar}\ \emph
  {et~al.}(2017{\natexlab{a}})\citenamefont {Benalcazar}, \citenamefont
  {Bernevig},\ and\ \citenamefont {Hughes}}]{Benalcazar17SCI}%
  \BibitemOpen
  \bibfield  {author} {\bibinfo {author} {\bibfnamefont {W.~A.}\ \bibnamefont
  {Benalcazar}}, \bibinfo {author} {\bibfnamefont {B.~A.}\ \bibnamefont
  {Bernevig}},\ and\ \bibinfo {author} {\bibfnamefont {T.~L.}\ \bibnamefont
  {Hughes}},\ }\bibfield  {title} {\bibinfo {title} {Quantized electric
  multipole insulators},\ }\href {https://doi.org/10.1126/science.aah6442}
  {\bibfield  {journal} {\bibinfo  {journal} {Science}\ }\textbf {\bibinfo
  {volume} {357}},\ \bibinfo {pages} {61} (\bibinfo {year}
  {2017}{\natexlab{a}})}\BibitemShut {NoStop}%
\bibitem [{\citenamefont {Benalcazar}\ \emph
  {et~al.}(2017{\natexlab{b}})\citenamefont {Benalcazar}, \citenamefont
  {Bernevig},\ and\ \citenamefont {Hughes}}]{Benalcazar17PRB}%
  \BibitemOpen
  \bibfield  {author} {\bibinfo {author} {\bibfnamefont {W.~A.}\ \bibnamefont
  {Benalcazar}}, \bibinfo {author} {\bibfnamefont {B.~A.}\ \bibnamefont
  {Bernevig}},\ and\ \bibinfo {author} {\bibfnamefont {T.~L.}\ \bibnamefont
  {Hughes}},\ }\bibfield  {title} {\bibinfo {title} {Electric multipole
  moments, topological multipole moment pumping, and chiral hinge states in
  crystalline insulators},\ }\href {https://doi.org/10.1103/PhysRevB.96.245115}
  {\bibfield  {journal} {\bibinfo  {journal} {Phys. Rev. B}\ }\textbf {\bibinfo
  {volume} {96}},\ \bibinfo {pages} {245115} (\bibinfo {year}
  {2017}{\natexlab{b}})}\BibitemShut {NoStop}%
\bibitem [{\citenamefont {Langbehn}\ \emph {et~al.}(2017)\citenamefont
  {Langbehn}, \citenamefont {Peng}, \citenamefont {Trifunovic}, \citenamefont
  {von Oppen},\ and\ \citenamefont {Brouwer}}]{Langbehn17PRL}%
  \BibitemOpen
  \bibfield  {author} {\bibinfo {author} {\bibfnamefont {J.}~\bibnamefont
  {Langbehn}}, \bibinfo {author} {\bibfnamefont {Y.}~\bibnamefont {Peng}},
  \bibinfo {author} {\bibfnamefont {L.}~\bibnamefont {Trifunovic}}, \bibinfo
  {author} {\bibfnamefont {F.}~\bibnamefont {von Oppen}},\ and\ \bibinfo
  {author} {\bibfnamefont {P.~W.}\ \bibnamefont {Brouwer}},\ }\bibfield
  {title} {\bibinfo {title} {Reflection-symmetric second-order topological
  insulators and superconductors},\ }\href
  {https://doi.org/10.1103/PhysRevLett.119.246401} {\bibfield  {journal}
  {\bibinfo  {journal} {Phys. Rev. Lett.}\ }\textbf {\bibinfo {volume} {119}},\
  \bibinfo {pages} {246401} (\bibinfo {year} {2017})}\BibitemShut {NoStop}%
\bibitem [{\citenamefont {Song}\ \emph {et~al.}(2017)\citenamefont {Song},
  \citenamefont {Fang},\ and\ \citenamefont {Fang}}]{SongZD17PRL}%
  \BibitemOpen
  \bibfield  {author} {\bibinfo {author} {\bibfnamefont {Z.}~\bibnamefont
  {Song}}, \bibinfo {author} {\bibfnamefont {Z.}~\bibnamefont {Fang}},\ and\
  \bibinfo {author} {\bibfnamefont {C.}~\bibnamefont {Fang}},\ }\bibfield
  {title} {\bibinfo {title} {$(d\ensuremath{-}2)$-dimensional edge states of
  rotation symmetry protected topological states},\ }\href
  {https://doi.org/10.1103/PhysRevLett.119.246402} {\bibfield  {journal}
  {\bibinfo  {journal} {Phys. Rev. Lett.}\ }\textbf {\bibinfo {volume} {119}},\
  \bibinfo {pages} {246402} (\bibinfo {year} {2017})}\BibitemShut {NoStop}%
\bibitem [{\citenamefont {Schindler}\ \emph {et~al.}(2018)\citenamefont
  {Schindler}, \citenamefont {Cook}, \citenamefont {Vergniory}, \citenamefont
  {Wang}, \citenamefont {Parkin}, \citenamefont {Bernevig},\ and\ \citenamefont
  {Neupert}}]{SchindlerF18SR}%
  \BibitemOpen
  \bibfield  {author} {\bibinfo {author} {\bibfnamefont {F.}~\bibnamefont
  {Schindler}}, \bibinfo {author} {\bibfnamefont {A.~M.}\ \bibnamefont {Cook}},
  \bibinfo {author} {\bibfnamefont {M.~G.}\ \bibnamefont {Vergniory}}, \bibinfo
  {author} {\bibfnamefont {Z.}~\bibnamefont {Wang}}, \bibinfo {author}
  {\bibfnamefont {S.~S.~P.}\ \bibnamefont {Parkin}}, \bibinfo {author}
  {\bibfnamefont {B.~A.}\ \bibnamefont {Bernevig}},\ and\ \bibinfo {author}
  {\bibfnamefont {T.}~\bibnamefont {Neupert}},\ }\bibfield  {title} {\bibinfo
  {title} {Higher-order topological insulators},\ }\href
  {https://doi.org/10.1126/sciadv.aat0346} {\bibfield  {journal} {\bibinfo
  {journal} {Sci. Adv.}\ }\textbf {\bibinfo {volume} {4}},\ \bibinfo {pages}
  {eaat0346} (\bibinfo {year} {2018})}\BibitemShut {NoStop}%
\bibitem [{\citenamefont {Fu}(2011)}]{FuL11PRL}%
  \BibitemOpen
  \bibfield  {author} {\bibinfo {author} {\bibfnamefont {L.}~\bibnamefont
  {Fu}},\ }\bibfield  {title} {\bibinfo {title} {Topological crystalline
  insulators},\ }\href {https://doi.org/10.1103/PhysRevLett.106.106802}
  {\bibfield  {journal} {\bibinfo  {journal} {Phys. Rev. Lett.}\ }\textbf
  {\bibinfo {volume} {106}},\ \bibinfo {pages} {106802} (\bibinfo {year}
  {2011})}\BibitemShut {NoStop}%
\bibitem [{\citenamefont {Slager}\ \emph {et~al.}(2012)\citenamefont {Slager},
  \citenamefont {Mesaros}, \citenamefont {Juri\ifmmode~\check{r}\else
  \v{c}\fi{}i\'c},\ and\ \citenamefont {Zaanen}}]{Slager12NTP}%
  \BibitemOpen
  \bibfield  {author} {\bibinfo {author} {\bibfnamefont {R.-J.}\ \bibnamefont
  {Slager}}, \bibinfo {author} {\bibfnamefont {A.}~\bibnamefont {Mesaros}},
  \bibinfo {author} {\bibfnamefont {V.}~\bibnamefont
  {Juri\ifmmode~\check{r}\else \v{c}\fi{}i\'c}},\ and\ \bibinfo {author}
  {\bibfnamefont {J.}~\bibnamefont {Zaanen}},\ }\bibfield  {title} {\bibinfo
  {title} {The space group classification of topological band-insulators},\
  }\href {https://doi.org/10.1038/nphys2513} {\bibfield  {journal} {\bibinfo
  {journal} {Nat. Phys.}\ }\textbf {\bibinfo {volume} {9}},\ \bibinfo {pages}
  {98} (\bibinfo {year} {2012})}\BibitemShut {NoStop}%
\bibitem [{\citenamefont {Kruthoff}\ \emph {et~al.}(2017)\citenamefont
  {Kruthoff}, \citenamefont {de~Boer}, \citenamefont {van Wezel}, \citenamefont
  {Kane},\ and\ \citenamefont {Slager}}]{KruthoffJ17PRX}%
  \BibitemOpen
  \bibfield  {author} {\bibinfo {author} {\bibfnamefont {J.}~\bibnamefont
  {Kruthoff}}, \bibinfo {author} {\bibfnamefont {J.}~\bibnamefont {de~Boer}},
  \bibinfo {author} {\bibfnamefont {J.}~\bibnamefont {van Wezel}}, \bibinfo
  {author} {\bibfnamefont {C.~L.}\ \bibnamefont {Kane}},\ and\ \bibinfo
  {author} {\bibfnamefont {R.-J.}\ \bibnamefont {Slager}},\ }\bibfield  {title}
  {\bibinfo {title} {Topological classification of crystalline insulators
  through band structure combinatorics},\ }\href
  {https://doi.org/10.1103/PhysRevX.7.041069} {\bibfield  {journal} {\bibinfo
  {journal} {Phys. Rev. X}\ }\textbf {\bibinfo {volume} {7}},\ \bibinfo {pages}
  {041069} (\bibinfo {year} {2017})}\BibitemShut {NoStop}%
\bibitem [{\citenamefont {Po}\ \emph {et~al.}(2017)\citenamefont {Po},
  \citenamefont {Vishwanath},\ and\ \citenamefont {Watanabe}}]{PoHC17NTC}%
  \BibitemOpen
  \bibfield  {author} {\bibinfo {author} {\bibfnamefont {H.~C.}\ \bibnamefont
  {Po}}, \bibinfo {author} {\bibfnamefont {A.}~\bibnamefont {Vishwanath}},\
  and\ \bibinfo {author} {\bibfnamefont {H.}~\bibnamefont {Watanabe}},\
  }\bibfield  {title} {\bibinfo {title} {Symmetry-based indicators of band
  topology in the 230 space groups},\ }\href
  {https://doi.org/10.1038/s41467-017-00133-2} {\bibfield  {journal} {\bibinfo
  {journal} {Nat. Commun.}\ }\textbf {\bibinfo {volume} {8}},\ \bibinfo {pages}
  {50} (\bibinfo {year} {2017})}\BibitemShut {NoStop}%
\bibitem [{\citenamefont {Bradlyn}\ \emph {et~al.}(2017)\citenamefont
  {Bradlyn}, \citenamefont {Elcoro}, \citenamefont {Cano}, \citenamefont
  {Vergniory}, \citenamefont {Wang}, \citenamefont {Felser}, \citenamefont
  {Aroyo},\ and\ \citenamefont {Bernevig}}]{BradlynB17NT}%
  \BibitemOpen
  \bibfield  {author} {\bibinfo {author} {\bibfnamefont {B.}~\bibnamefont
  {Bradlyn}}, \bibinfo {author} {\bibfnamefont {L.}~\bibnamefont {Elcoro}},
  \bibinfo {author} {\bibfnamefont {J.}~\bibnamefont {Cano}}, \bibinfo {author}
  {\bibfnamefont {M.~G.}\ \bibnamefont {Vergniory}}, \bibinfo {author}
  {\bibfnamefont {Z.}~\bibnamefont {Wang}}, \bibinfo {author} {\bibfnamefont
  {C.}~\bibnamefont {Felser}}, \bibinfo {author} {\bibfnamefont {M.~I.}\
  \bibnamefont {Aroyo}},\ and\ \bibinfo {author} {\bibfnamefont {B.~A.}\
  \bibnamefont {Bernevig}},\ }\bibfield  {title} {\bibinfo {title} {Topological
  quantum chemistry},\ }\href {https://doi.org/10.1038/nature23268} {\bibfield
  {journal} {\bibinfo  {journal} {Nature}\ }\textbf {\bibinfo {volume} {547}},\
  \bibinfo {pages} {298} (\bibinfo {year} {2017})}\BibitemShut {NoStop}%
\bibitem [{\citenamefont {Nayak}\ \emph {et~al.}(2008)\citenamefont {Nayak},
  \citenamefont {Simon}, \citenamefont {Stern}, \citenamefont {Freedman},\ and\
  \citenamefont {Das~Sarma}}]{Nayak08RMP}%
  \BibitemOpen
  \bibfield  {author} {\bibinfo {author} {\bibfnamefont {C.}~\bibnamefont
  {Nayak}}, \bibinfo {author} {\bibfnamefont {S.~H.}\ \bibnamefont {Simon}},
  \bibinfo {author} {\bibfnamefont {A.}~\bibnamefont {Stern}}, \bibinfo
  {author} {\bibfnamefont {M.}~\bibnamefont {Freedman}},\ and\ \bibinfo
  {author} {\bibfnamefont {S.}~\bibnamefont {Das~Sarma}},\ }\bibfield  {title}
  {\bibinfo {title} {Non-abelian anyons and topological quantum computation},\
  }\href {https://doi.org/10.1103/RevModPhys.80.1083} {\bibfield  {journal}
  {\bibinfo  {journal} {Rev. Mod. Phys.}\ }\textbf {\bibinfo {volume} {80}},\
  \bibinfo {pages} {1083} (\bibinfo {year} {2008})}\BibitemShut {NoStop}%
\bibitem [{\citenamefont {Anderson}(1958)}]{Anderson58PR}%
  \BibitemOpen
  \bibfield  {author} {\bibinfo {author} {\bibfnamefont {P.~W.}\ \bibnamefont
  {Anderson}},\ }\bibfield  {title} {\bibinfo {title} {Absence of diffusion in
  certain random lattices},\ }\href {https://doi.org/10.1103/PhysRev.109.1492}
  {\bibfield  {journal} {\bibinfo  {journal} {Phys. Rev.}\ }\textbf {\bibinfo
  {volume} {109}},\ \bibinfo {pages} {1492} (\bibinfo {year}
  {1958})}\BibitemShut {NoStop}%
\bibitem [{\citenamefont {Abrahams}\ \emph {et~al.}(1979)\citenamefont
  {Abrahams}, \citenamefont {Anderson}, \citenamefont {Licciardello},\ and\
  \citenamefont {Ramakrishnan}}]{AbrahamsE79PRL}%
  \BibitemOpen
  \bibfield  {author} {\bibinfo {author} {\bibfnamefont {E.}~\bibnamefont
  {Abrahams}}, \bibinfo {author} {\bibfnamefont {P.~W.}\ \bibnamefont
  {Anderson}}, \bibinfo {author} {\bibfnamefont {D.~C.}\ \bibnamefont
  {Licciardello}},\ and\ \bibinfo {author} {\bibfnamefont {T.~V.}\ \bibnamefont
  {Ramakrishnan}},\ }\bibfield  {title} {\bibinfo {title} {Scaling theory of
  localization: Absence of quantum diffusion in two dimensions},\ }\href
  {https://doi.org/10.1103/PhysRevLett.42.673} {\bibfield  {journal} {\bibinfo
  {journal} {Phys. Rev. Lett.}\ }\textbf {\bibinfo {volume} {42}},\ \bibinfo
  {pages} {673} (\bibinfo {year} {1979})}\BibitemShut {NoStop}%
\bibitem [{\citenamefont {Evers}\ and\ \citenamefont
  {Mirlin}(2008)}]{FerdinandE08RMP}%
  \BibitemOpen
  \bibfield  {author} {\bibinfo {author} {\bibfnamefont {F.}~\bibnamefont
  {Evers}}\ and\ \bibinfo {author} {\bibfnamefont {A.~D.}\ \bibnamefont
  {Mirlin}},\ }\bibfield  {title} {\bibinfo {title} {Anderson transitions},\
  }\href {https://doi.org/10.1103/RevModPhys.80.1355} {\bibfield  {journal}
  {\bibinfo  {journal} {Rev. Mod. Phys.}\ }\textbf {\bibinfo {volume} {80}},\
  \bibinfo {pages} {1355} (\bibinfo {year} {2008})}\BibitemShut {NoStop}%
\bibitem [{\citenamefont {Dunlap}\ \emph {et~al.}(1990)\citenamefont {Dunlap},
  \citenamefont {Wu},\ and\ \citenamefont {Phillips}}]{DunlapDH90PRL}%
  \BibitemOpen
  \bibfield  {author} {\bibinfo {author} {\bibfnamefont {D.~H.}\ \bibnamefont
  {Dunlap}}, \bibinfo {author} {\bibfnamefont {H.-L.}\ \bibnamefont {Wu}},\
  and\ \bibinfo {author} {\bibfnamefont {P.~W.}\ \bibnamefont {Phillips}},\
  }\bibfield  {title} {\bibinfo {title} {Absence of localization in a
  random-dimer model},\ }\href {https://doi.org/10.1103/PhysRevLett.65.88}
  {\bibfield  {journal} {\bibinfo  {journal} {Phys. Rev. Lett.}\ }\textbf
  {\bibinfo {volume} {65}},\ \bibinfo {pages} {88} (\bibinfo {year}
  {1990})}\BibitemShut {NoStop}%
\bibitem [{\citenamefont {Phillips}\ and\ \citenamefont
  {Wu}(1991)}]{PhillipsP91SCI}%
  \BibitemOpen
  \bibfield  {author} {\bibinfo {author} {\bibfnamefont {P.}~\bibnamefont
  {Phillips}}\ and\ \bibinfo {author} {\bibfnamefont {H.~L.}\ \bibnamefont
  {Wu}},\ }\bibfield  {title} {\bibinfo {title} {Localization and its absence:
  A new metallic state for conducting polymers},\ }\href
  {https://doi.org/10.1126/science.252.5014.1805} {\bibfield  {journal}
  {\bibinfo  {journal} {Science}\ }\textbf {\bibinfo {volume} {252}},\ \bibinfo
  {pages} {1805} (\bibinfo {year} {1991})}\BibitemShut {NoStop}%
\bibitem [{\citenamefont {Wu}\ and\ \citenamefont
  {Phillips}(1991)}]{WuHL91PRL}%
  \BibitemOpen
  \bibfield  {author} {\bibinfo {author} {\bibfnamefont {H.-L.}\ \bibnamefont
  {Wu}}\ and\ \bibinfo {author} {\bibfnamefont {P.}~\bibnamefont {Phillips}},\
  }\bibfield  {title} {\bibinfo {title} {Polyaniline is a random-dimer model: A
  new transport mechanism for conducting polymers},\ }\href
  {https://doi.org/10.1103/PhysRevLett.66.1366} {\bibfield  {journal} {\bibinfo
   {journal} {Phys. Rev. Lett.}\ }\textbf {\bibinfo {volume} {66}},\ \bibinfo
  {pages} {1366} (\bibinfo {year} {1991})}\BibitemShut {NoStop}%
\bibitem [{\citenamefont {Bellani}\ \emph {et~al.}(1999)\citenamefont
  {Bellani}, \citenamefont {Diez}, \citenamefont {Hey}, \citenamefont {Toni},
  \citenamefont {Tarricone}, \citenamefont {Parravicini}, \citenamefont
  {Dom\'{\i}nguez-Adame},\ and\ \citenamefont
  {G\'omez-Alcal\'a}}]{BellaniV99PRL}%
  \BibitemOpen
  \bibfield  {author} {\bibinfo {author} {\bibfnamefont {V.}~\bibnamefont
  {Bellani}}, \bibinfo {author} {\bibfnamefont {E.}~\bibnamefont {Diez}},
  \bibinfo {author} {\bibfnamefont {R.}~\bibnamefont {Hey}}, \bibinfo {author}
  {\bibfnamefont {L.}~\bibnamefont {Toni}}, \bibinfo {author} {\bibfnamefont
  {L.}~\bibnamefont {Tarricone}}, \bibinfo {author} {\bibfnamefont {G.~B.}\
  \bibnamefont {Parravicini}}, \bibinfo {author} {\bibfnamefont
  {F.}~\bibnamefont {Dom\'{\i}nguez-Adame}},\ and\ \bibinfo {author}
  {\bibfnamefont {R.}~\bibnamefont {G\'omez-Alcal\'a}},\ }\bibfield  {title}
  {\bibinfo {title} {Experimental evidence of delocalized states in random
  dimer superlattices},\ }\href {https://doi.org/10.1103/PhysRevLett.82.2159}
  {\bibfield  {journal} {\bibinfo  {journal} {Phys. Rev. Lett.}\ }\textbf
  {\bibinfo {volume} {82}},\ \bibinfo {pages} {2159} (\bibinfo {year}
  {1999})}\BibitemShut {NoStop}%
\bibitem [{\citenamefont {Naether}\ \emph {et~al.}(2013)\citenamefont
  {Naether}, \citenamefont {St\"utzer}, \citenamefont {Vicencio}, \citenamefont
  {Molina}, \citenamefont {T\"unnermann}, \citenamefont {Nolte}, \citenamefont
  {Kottos}, \citenamefont {Christodoulides},\ and\ \citenamefont
  {Szameit}}]{NaetherN13NJP}%
  \BibitemOpen
  \bibfield  {author} {\bibinfo {author} {\bibfnamefont {U.}~\bibnamefont
  {Naether}}, \bibinfo {author} {\bibfnamefont {S.}~\bibnamefont {St\"utzer}},
  \bibinfo {author} {\bibfnamefont {R.~A.}\ \bibnamefont {Vicencio}}, \bibinfo
  {author} {\bibfnamefont {M.~I.}\ \bibnamefont {Molina}}, \bibinfo {author}
  {\bibfnamefont {A.}~\bibnamefont {T\"unnermann}}, \bibinfo {author}
  {\bibfnamefont {S.}~\bibnamefont {Nolte}}, \bibinfo {author} {\bibfnamefont
  {T.}~\bibnamefont {Kottos}}, \bibinfo {author} {\bibfnamefont {D.~N.}\
  \bibnamefont {Christodoulides}},\ and\ \bibinfo {author} {\bibfnamefont
  {A.}~\bibnamefont {Szameit}},\ }\bibfield  {title} {\bibinfo {title}
  {Experimental observation of superdiffusive transport in random dimer
  lattices},\ }\href {https://doi.org/10.1088/1367-2630/15/1/013045} {\bibfield
   {journal} {\bibinfo  {journal} {New J. Phys.}\ }\textbf {\bibinfo {volume}
  {15}},\ \bibinfo {pages} {013045} (\bibinfo {year} {2013})}\BibitemShut
  {NoStop}%
\bibitem [{\citenamefont {Mondragon-Shem}\ \emph {et~al.}(2014)\citenamefont
  {Mondragon-Shem}, \citenamefont {Hughes}, \citenamefont {Song},\ and\
  \citenamefont {Prodan}}]{MondragonShem14PRL}%
  \BibitemOpen
  \bibfield  {author} {\bibinfo {author} {\bibfnamefont {I.}~\bibnamefont
  {Mondragon-Shem}}, \bibinfo {author} {\bibfnamefont {T.~L.}\ \bibnamefont
  {Hughes}}, \bibinfo {author} {\bibfnamefont {J.}~\bibnamefont {Song}},\ and\
  \bibinfo {author} {\bibfnamefont {E.}~\bibnamefont {Prodan}},\ }\bibfield
  {title} {\bibinfo {title} {Topological criticality in the chiral-symmetric
AIII class at strong disorder},\ }\href
  {https://doi.org/10.1103/PhysRevLett.113.046802} {\bibfield  {journal}
  {\bibinfo  {journal} {Phys. Rev. Lett.}\ }\textbf {\bibinfo {volume} {113}},\
  \bibinfo {pages} {046802} (\bibinfo {year} {2014})}\BibitemShut {NoStop}%
\bibitem [{\citenamefont {Altland}\ \emph {et~al.}(2014)\citenamefont
  {Altland}, \citenamefont {Bagrets}, \citenamefont {Fritz}, \citenamefont
  {Kamenev},\ and\ \citenamefont {Schmiedt}}]{AltlandA14PRL2}%
  \BibitemOpen
  \bibfield  {author} {\bibinfo {author} {\bibfnamefont {A.}~\bibnamefont
  {Altland}}, \bibinfo {author} {\bibfnamefont {D.}~\bibnamefont {Bagrets}},
  \bibinfo {author} {\bibfnamefont {L.}~\bibnamefont {Fritz}}, \bibinfo
  {author} {\bibfnamefont {A.}~\bibnamefont {Kamenev}},\ and\ \bibinfo {author}
  {\bibfnamefont {H.}~\bibnamefont {Schmiedt}},\ }\bibfield  {title} {\bibinfo
  {title} {Quantum criticality of quasi-one-dimensional topological anderson
  insulators},\ }\href {https://doi.org/10.1103/PhysRevLett.112.206602}
  {\bibfield  {journal} {\bibinfo  {journal} {Phys. Rev. Lett.}\ }\textbf
  {\bibinfo {volume} {112}},\ \bibinfo {pages} {206602} (\bibinfo {year}
  {2014})}\BibitemShut {NoStop}%
\bibitem [{\citenamefont {Girschik}\ \emph {et~al.}(2013)\citenamefont
  {Girschik}, \citenamefont {Libisch},\ and\ \citenamefont
  {Rotter}}]{GirschikA13PRB}%
  \BibitemOpen
  \bibfield  {author} {\bibinfo {author} {\bibfnamefont {A.}~\bibnamefont
  {Girschik}}, \bibinfo {author} {\bibfnamefont {F.}~\bibnamefont {Libisch}},\
  and\ \bibinfo {author} {\bibfnamefont {S.}~\bibnamefont {Rotter}},\
  }\bibfield  {title} {\bibinfo {title} {Topological insulator in the presence
  of spatially correlated disorder},\ }\href
  {https://doi.org/10.1103/PhysRevB.88.014201} {\bibfield  {journal} {\bibinfo
  {journal} {Phys. Rev. B}\ }\textbf {\bibinfo {volume} {88}},\ \bibinfo
  {pages} {014201} (\bibinfo {year} {2013})}\BibitemShut {NoStop}%
\bibitem [{\citenamefont {Wauters}\ \emph {et~al.}(2019)\citenamefont
  {Wauters}, \citenamefont {Russomanno}, \citenamefont {Citro}, \citenamefont
  {Santoro},\ and\ \citenamefont {Privitera}}]{WautersMM19PRL}%
  \BibitemOpen
  \bibfield  {author} {\bibinfo {author} {\bibfnamefont {M.~M.}\ \bibnamefont
  {Wauters}}, \bibinfo {author} {\bibfnamefont {A.}~\bibnamefont {Russomanno}},
  \bibinfo {author} {\bibfnamefont {R.}~\bibnamefont {Citro}}, \bibinfo
  {author} {\bibfnamefont {G.~E.}\ \bibnamefont {Santoro}},\ and\ \bibinfo
  {author} {\bibfnamefont {L.}~\bibnamefont {Privitera}},\ }\bibfield  {title}
  {\bibinfo {title} {Localization, topology, and quantized transport in
  disordered Floquet systems},\ }\href
  {https://doi.org/10.1103/PhysRevLett.123.266601} {\bibfield  {journal}
  {\bibinfo  {journal} {Phys. Rev. Lett.}\ }\textbf {\bibinfo {volume} {123}},\
  \bibinfo {pages} {266601} (\bibinfo {year} {2019})}\BibitemShut {NoStop}%
\bibitem [{\citenamefont {Hayward}\ \emph {et~al.}(2021)\citenamefont
  {Hayward}, \citenamefont {Bertok}, \citenamefont {Schneider},\ and\
  \citenamefont {Heidrich-Meisner}}]{HaywardALC21PRA}%
  \BibitemOpen
  \bibfield  {author} {\bibinfo {author} {\bibfnamefont {A.~L.~C.}\
  \bibnamefont {Hayward}}, \bibinfo {author} {\bibfnamefont {E.}~\bibnamefont
  {Bertok}}, \bibinfo {author} {\bibfnamefont {U.}~\bibnamefont {Schneider}},\
  and\ \bibinfo {author} {\bibfnamefont {F.}~\bibnamefont {Heidrich-Meisner}},\
  }\bibfield  {title} {\bibinfo {title} {Effect of disorder on topological
  charge pumping in the Rice-Mele model},\ }\href
  {https://doi.org/10.1103/PhysRevA.103.043310} {\bibfield  {journal} {\bibinfo
   {journal} {Phys. Rev. A}\ }\textbf {\bibinfo {volume} {103}},\ \bibinfo
  {pages} {043310} (\bibinfo {year} {2021})}\BibitemShut {NoStop}%
\bibitem [{\citenamefont {Li}\ \emph {et~al.}(2009)\citenamefont {Li},
  \citenamefont {Chu}, \citenamefont {Jain},\ and\ \citenamefont
  {Shen}}]{LiJ09PRL}%
  \BibitemOpen
  \bibfield  {author} {\bibinfo {author} {\bibfnamefont {J.}~\bibnamefont
  {Li}}, \bibinfo {author} {\bibfnamefont {R.-L.}\ \bibnamefont {Chu}},
  \bibinfo {author} {\bibfnamefont {J.~K.}\ \bibnamefont {Jain}},\ and\
  \bibinfo {author} {\bibfnamefont {S.-Q.}\ \bibnamefont {Shen}},\ }\bibfield
  {title} {\bibinfo {title} {Topological Anderson insulator},\ }\href
  {https://doi.org/10.1103/PhysRevLett.102.136806} {\bibfield  {journal}
  {\bibinfo  {journal} {Phys. Rev. Lett.}\ }\textbf {\bibinfo {volume} {102}},\
  \bibinfo {pages} {136806} (\bibinfo {year} {2009})}\BibitemShut {NoStop}%
\bibitem [{\citenamefont {Groth}\ \emph {et~al.}(2009)\citenamefont {Groth},
  \citenamefont {Wimmer}, \citenamefont {Akhmerov}, \citenamefont
  {Tworzyd\l{}o},\ and\ \citenamefont {Beenakker}}]{GrothCW09PRL}%
  \BibitemOpen
  \bibfield  {author} {\bibinfo {author} {\bibfnamefont {C.~W.}\ \bibnamefont
  {Groth}}, \bibinfo {author} {\bibfnamefont {M.}~\bibnamefont {Wimmer}},
  \bibinfo {author} {\bibfnamefont {A.~R.}\ \bibnamefont {Akhmerov}}, \bibinfo
  {author} {\bibfnamefont {J.}~\bibnamefont {Tworzyd\l{}o}},\ and\ \bibinfo
  {author} {\bibfnamefont {C.~W.~J.}\ \bibnamefont {Beenakker}},\ }\bibfield
  {title} {\bibinfo {title} {Theory of the topological Anderson insulator},\
  }\href {https://doi.org/10.1103/PhysRevLett.103.196805} {\bibfield  {journal}
  {\bibinfo  {journal} {Phys. Rev. Lett.}\ }\textbf {\bibinfo {volume} {103}},\
  \bibinfo {pages} {196805} (\bibinfo {year} {2009})}\BibitemShut {NoStop}%
\bibitem [{\citenamefont {Guo}\ \emph {et~al.}(2010)\citenamefont {Guo},
  \citenamefont {Rosenberg}, \citenamefont {Refael},\ and\ \citenamefont
  {Franz}}]{GuoHM10PRL}%
  \BibitemOpen
  \bibfield  {author} {\bibinfo {author} {\bibfnamefont {H.-M.}\ \bibnamefont
  {Guo}}, \bibinfo {author} {\bibfnamefont {G.}~\bibnamefont {Rosenberg}},
  \bibinfo {author} {\bibfnamefont {G.}~\bibnamefont {Refael}},\ and\ \bibinfo
  {author} {\bibfnamefont {M.}~\bibnamefont {Franz}},\ }\bibfield  {title}
  {\bibinfo {title} {Topological Anderson insulator in three dimensions},\
  }\href {https://doi.org/10.1103/PhysRevLett.105.216601} {\bibfield  {journal}
  {\bibinfo  {journal} {Phys. Rev. Lett.}\ }\textbf {\bibinfo {volume} {105}},\
  \bibinfo {pages} {216601} (\bibinfo {year} {2010})}\BibitemShut {NoStop}%
\bibitem [{\citenamefont {Meier}\ \emph {et~al.}(2018)\citenamefont {Meier},
  \citenamefont {An}, \citenamefont {Dauphin}, \citenamefont {Maffei},
  \citenamefont {Massignan}, \citenamefont {Hughes},\ and\ \citenamefont
  {Gadway}}]{MeierEJ18SCI}%
  \BibitemOpen
  \bibfield  {author} {\bibinfo {author} {\bibfnamefont {E.~J.}\ \bibnamefont
  {Meier}}, \bibinfo {author} {\bibfnamefont {F.~A.}\ \bibnamefont {An}},
  \bibinfo {author} {\bibfnamefont {A.}~\bibnamefont {Dauphin}}, \bibinfo
  {author} {\bibfnamefont {M.}~\bibnamefont {Maffei}}, \bibinfo {author}
  {\bibfnamefont {P.}~\bibnamefont {Massignan}}, \bibinfo {author}
  {\bibfnamefont {T.~L.}\ \bibnamefont {Hughes}},\ and\ \bibinfo {author}
  {\bibfnamefont {B.}~\bibnamefont {Gadway}},\ }\bibfield  {title} {\bibinfo
  {title} {Observation of the topological Anderson insulator in disordered
  atomic wires},\ }\href {https://doi.org/10.1126/science.aat3406} {\bibfield
  {journal} {\bibinfo  {journal} {Science}\ }\textbf {\bibinfo {volume}
  {362}},\ \bibinfo {pages} {929} (\bibinfo {year} {2018})}\BibitemShut
  {NoStop}%
\bibitem [{\citenamefont {Zhang}\ \emph {et~al.}(2020)\citenamefont {Zhang},
  \citenamefont {Tang}, \citenamefont {Lang}, \citenamefont {Yan},\ and\
  \citenamefont {Zhu}}]{ZhangDW20SC}%
  \BibitemOpen
  \bibfield  {author} {\bibinfo {author} {\bibfnamefont {D.-W.}\ \bibnamefont
  {Zhang}}, \bibinfo {author} {\bibfnamefont {L.-Z.}\ \bibnamefont {Tang}},
  \bibinfo {author} {\bibfnamefont {L.-J.}\ \bibnamefont {Lang}}, \bibinfo
  {author} {\bibfnamefont {H.}~\bibnamefont {Yan}},\ and\ \bibinfo {author}
  {\bibfnamefont {S.-L.}\ \bibnamefont {Zhu}},\ }\bibfield  {title} {\bibinfo
  {title} {Non-hermitian topological Anderson insulators},\ }\href
  {https://doi.org/10.1007/s11433-020-1521-9} {\bibfield  {journal} {\bibinfo
  {journal} {Sci. China Phys. Mech. Astron.}\ }\textbf {\bibinfo {volume}
  {63}},\ \bibinfo {pages} {267062} (\bibinfo {year} {2020})}\BibitemShut
  {NoStop}%
\bibitem [{\citenamefont {Luo}\ and\ \citenamefont
  {Zhang}(2019)}]{LuoXW19arXiv}%
  \BibitemOpen
  \bibfield  {author} {\bibinfo {author} {\bibfnamefont {X.-W.}\ \bibnamefont
  {Luo}}\ and\ \bibinfo {author} {\bibfnamefont {C.}~\bibnamefont {Zhang}},\
  }\bibfield  {title} {\bibinfo {title} {Non-Hermitian disorder-induced
  topological insulators},\ }\href {https://arxiv.org/abs/1912.10652}
  {\bibfield  {journal} {\bibinfo  {journal} {arXiv:1912.10652}\ } (\bibinfo
  {year} {2019})}
  \BibitemShut {NoStop}%
\bibitem [{\citenamefont {Su}\ \emph {et~al.}(1979)\citenamefont {Su},
  \citenamefont {Schrieffer},\ and\ \citenamefont {Heeger}}]{SuWP79PRL}%
  \BibitemOpen
  \bibfield  {author} {\bibinfo {author} {\bibfnamefont {W.~P.}\ \bibnamefont
  {Su}}, \bibinfo {author} {\bibfnamefont {J.~R.}\ \bibnamefont {Schrieffer}},\
  and\ \bibinfo {author} {\bibfnamefont {A.~J.}\ \bibnamefont {Heeger}},\
  }\bibfield  {title} {\bibinfo {title} {Solitons in polyacetylene},\ }\href
  {https://doi.org/10.1103/PhysRevLett.42.1698} {\bibfield  {journal} {\bibinfo
   {journal} {Phys. Rev. Lett.}\ }\textbf {\bibinfo {volume} {42}},\ \bibinfo
  {pages} {1698} (\bibinfo {year} {1979})}\BibitemShut {NoStop}%
\bibitem [{\citenamefont {Liu}\ and\ \citenamefont {Guo}(2018)}]{LiuT18PLA}%
  \BibitemOpen
  \bibfield  {author} {\bibinfo {author} {\bibfnamefont {T.}~\bibnamefont
  {Liu}}\ and\ \bibinfo {author} {\bibfnamefont {H.}~\bibnamefont {Guo}},\
  }\bibfield  {title} {\bibinfo {title} {Topological phase transition in the
  quasiperiodic disordered Su{\textendash}Schriffer{\textendash}Heeger chain},\
  }\href {https://doi.org/10.1016/j.physleta.2018.09.023} {\bibfield  {journal}
  {\bibinfo  {journal} {Phys. Lett. A}\ }\textbf {\bibinfo {volume} {382}},\
  \bibinfo {pages} {3287} (\bibinfo {year} {2018})}\BibitemShut {NoStop}%
\bibitem [{\citenamefont {Longhi}(2020)}]{LonghiS20OL}%
  \BibitemOpen
  \bibfield  {author} {\bibinfo {author} {\bibfnamefont {S.}~\bibnamefont
  {Longhi}},\ }\bibfield  {title} {\bibinfo {title} {Topological Anderson phase
  in quasi-periodic waveguide lattices},\ }\href
  {https://doi.org/10.1364/OL.399742} {\bibfield  {journal} {\bibinfo
  {journal} {Opt. Lett.}\ }\textbf {\bibinfo {volume} {45}},\ \bibinfo {pages}
  {4036} (\bibinfo {year} {2020})}\BibitemShut {NoStop}%
\bibitem [{\citenamefont {P\'erez-Gonz\'alez}\ \emph
  {et~al.}(2019)\citenamefont {P\'erez-Gonz\'alez}, \citenamefont {Bello},
  \citenamefont {G\'omez-Le\'on},\ and\ \citenamefont
  {Platero}}]{BeatrizP19PRB}%
  \BibitemOpen
  \bibfield  {author} {\bibinfo {author} {\bibfnamefont {B.}~\bibnamefont
  {P\'erez-Gonz\'alez}}, \bibinfo {author} {\bibfnamefont {M.}~\bibnamefont
  {Bello}}, \bibinfo {author} {\bibfnamefont {A.}~\bibnamefont
  {G\'omez-Le\'on}},\ and\ \bibinfo {author} {\bibfnamefont {G.}~\bibnamefont
  {Platero}},\ }\bibfield  {title} {\bibinfo {title} {Interplay between
  long-range hopping and disorder in topological systems},\ }\href
  {https://doi.org/10.1103/PhysRevB.99.035146} {\bibfield  {journal} {\bibinfo
  {journal} {Phys. Rev. B}\ }\textbf {\bibinfo {volume} {99}},\ \bibinfo
  {pages} {035146} (\bibinfo {year} {2019})}\BibitemShut {NoStop}%
\bibitem [{\citenamefont {Scollon}\ and\ \citenamefont
  {Kennett}(2020)}]{ScollonM20PRB}%
  \BibitemOpen
  \bibfield  {author} {\bibinfo {author} {\bibfnamefont {M.}~\bibnamefont
  {Scollon}}\ and\ \bibinfo {author} {\bibfnamefont {M.~P.}\ \bibnamefont
  {Kennett}},\ }\bibfield  {title} {\bibinfo {title} {Persistence of chirality
  in the Su-Schrieffer-Heeger model in the presence of on-site disorder},\
  }\href {https://doi.org/10.1103/PhysRevB.101.144204} {\bibfield  {journal}
  {\bibinfo  {journal} {Phys. Rev. B}\ }\textbf {\bibinfo {volume} {101}},\
  \bibinfo {pages} {144204} (\bibinfo {year} {2020})}\BibitemShut {NoStop}%
\bibitem [{\citenamefont {Roy}\ \emph {et~al.}(2021)\citenamefont {Roy},
  \citenamefont {Mishra}, \citenamefont {Tanatar},\ and\ \citenamefont
  {Basu}}]{ShilpiR21PRL}%
  \BibitemOpen
  \bibfield  {author} {\bibinfo {author} {\bibfnamefont {S.}~\bibnamefont
  {Roy}}, \bibinfo {author} {\bibfnamefont {T.}~\bibnamefont {Mishra}},
  \bibinfo {author} {\bibfnamefont {B.}~\bibnamefont {Tanatar}},\ and\ \bibinfo
  {author} {\bibfnamefont {S.}~\bibnamefont {Basu}},\ }\bibfield  {title}
  {\bibinfo {title} {Reentrant localization transition in a quasiperiodic
  chain},\ }\href {https://doi.org/10.1103/PhysRevLett.126.106803} {\bibfield
  {journal} {\bibinfo  {journal} {Phys. Rev. Lett.}\ }\textbf {\bibinfo
  {volume} {126}},\ \bibinfo {pages} {106803} (\bibinfo {year}
  {2021})}\BibitemShut {NoStop}%
\bibitem [{\citenamefont {Shen}(2012)}]{ShenSQ12book}%
  \BibitemOpen
  \bibfield  {author} {\bibinfo {author} {\bibfnamefont {S.-Q.}\ \bibnamefont
  {Shen}},\ }\href {https://doi.org/10.1007/978-3-642-32858-9} {\emph {\bibinfo
  {title} {Topological Insulators: Dirac Equation in Condensed Matters}}}\
  (\bibinfo  {publisher} {Springer},\ \bibinfo {year} {2012})\BibitemShut
  {NoStop}%
\bibitem [{\citenamefont {Asb{\'{o}}th}\ \emph {et~al.}(2016)\citenamefont
  {Asb{\'{o}}th}, \citenamefont {Oroszl{\'{a}}ny},\ and\ \citenamefont
  {P{\'{a}}lyi}}]{AsbothJ16}%
  \BibitemOpen
  \bibfield  {author} {\bibinfo {author} {\bibfnamefont {J.~K.}\ \bibnamefont
  {Asb{\'{o}}th}}, \bibinfo {author} {\bibfnamefont {L.}~\bibnamefont
  {Oroszl{\'{a}}ny}},\ and\ \bibinfo {author} {\bibfnamefont {A.}~\bibnamefont
  {P{\'{a}}lyi}},\ }\href {https://doi.org/10.1007/978-3-319-25607-8} {\emph
  {\bibinfo {title} {A Short Course on Topological Insulators}}}\ (\bibinfo
  {publisher} {Springer},\ \bibinfo {year} {2016})\BibitemShut {NoStop}%
\bibitem [{\citenamefont {Resta}(1998)}]{Resta98PRL}%
  \BibitemOpen
  \bibfield  {author} {\bibinfo {author} {\bibfnamefont {R.}~\bibnamefont
  {Resta}},\ }\bibfield  {title} {\bibinfo {title} {Quantum-mechanical position
  operator in extended systems},\ }\href
  {https://doi.org/10.1103/PhysRevLett.80.1800} {\bibfield  {journal} {\bibinfo
   {journal} {Phys. Rev. Lett.}\ }\textbf {\bibinfo {volume} {80}},\ \bibinfo
  {pages} {1800} (\bibinfo {year} {1998})}\BibitemShut {NoStop}%
\bibitem [{\citenamefont {Rhim}\ \emph {et~al.}(2018)\citenamefont {Rhim},
  \citenamefont {Bardarson},\ and\ \citenamefont {Slager}}]{RhimJW18PRB}%
  \BibitemOpen
  \bibfield  {author} {\bibinfo {author} {\bibfnamefont {J.-W.}\ \bibnamefont
  {Rhim}}, \bibinfo {author} {\bibfnamefont {J.~H.}\ \bibnamefont
  {Bardarson}},\ and\ \bibinfo {author} {\bibfnamefont {R.-J.}\ \bibnamefont
  {Slager}},\ }\bibfield  {title} {\bibinfo {title} {Unified bulk-boundary
  correspondence for band insulators},\ }\href
  {https://doi.org/10.1103/PhysRevB.97.115143} {\bibfield  {journal} {\bibinfo
  {journal} {Phys. Rev. B}\ }\textbf {\bibinfo {volume} {97}},\ \bibinfo
  {pages} {115143} (\bibinfo {year} {2018})}\BibitemShut {NoStop}%
\bibitem [{\citenamefont {Li}\ and\ \citenamefont
  {Das~Sarma}(2020)}]{LiX20PRB}%
  \BibitemOpen
  \bibfield  {author} {\bibinfo {author} {\bibfnamefont {X.}~\bibnamefont
  {Li}}\ and\ \bibinfo {author} {\bibfnamefont {S.}~\bibnamefont {Das~Sarma}},\
  }\bibfield  {title} {\bibinfo {title} {Mobility edge and intermediate phase
  in one-dimensional incommensurate lattice potentials},\ }\href
  {https://doi.org/10.1103/PhysRevB.101.064203} {\bibfield  {journal} {\bibinfo
   {journal} {Phys. Rev. B}\ }\textbf {\bibinfo {volume} {101}},\ \bibinfo
  {pages} {064203} (\bibinfo {year} {2020})}\BibitemShut {NoStop}%
\bibitem [{\citenamefont {Groth}\ \emph {et~al.}(2014)\citenamefont {Groth},
  \citenamefont {Wimmer}, \citenamefont {Akhmerov},\ and\ \citenamefont
  {Waintal}}]{GrothCW14NJP}%
  \BibitemOpen
  \bibfield  {author} {\bibinfo {author} {\bibfnamefont {C.~W.}\ \bibnamefont
  {Groth}}, \bibinfo {author} {\bibfnamefont {M.}~\bibnamefont {Wimmer}},
  \bibinfo {author} {\bibfnamefont {A.~R.}\ \bibnamefont {Akhmerov}},\ and\
  \bibinfo {author} {\bibfnamefont {X.}~\bibnamefont {Waintal}},\ }\bibfield
  {title} {\bibinfo {title} {Kwant: a software package for quantum transport},\
  }\href {https://doi.org/10.1088/1367-2630/16/6/063065} {\bibfield  {journal}
  {\bibinfo  {journal} {New J. Phys.}\ }\textbf {\bibinfo {volume} {16}},\
  \bibinfo {pages} {063065} (\bibinfo {year} {2014})}\BibitemShut {NoStop}%
\bibitem [{\citenamefont {Kawabata}\ and\ \citenamefont
  {Ryu}(2021)}]{KawabataK21PRL}%
  \BibitemOpen
  \bibfield  {author} {\bibinfo {author} {\bibfnamefont {K.}~\bibnamefont
  {Kawabata}}\ and\ \bibinfo {author} {\bibfnamefont {S.}~\bibnamefont {Ryu}},\
  }\bibfield  {title} {\bibinfo {title} {Nonunitary scaling theory of
  non-Hermitian localization},\ }\href
  {https://doi.org/10.1103/PhysRevLett.126.166801} {\bibfield  {journal}
  {\bibinfo  {journal} {Phys. Rev. Lett.}\ }\textbf {\bibinfo {volume} {126}},\
  \bibinfo {pages} {166801} (\bibinfo {year} {2021})}\BibitemShut {NoStop}%
   \bibitem [{\citenamefont {Liu}\ \emph {et~al.}(2022)\citenamefont {Liu},
  \citenamefont {Zhang}, \citenamefont {Tang},\ and\ \citenamefont
  {Zhang}}]{LiuSH22PLA}%
  \BibitemOpen
  \bibfield  {author} {\bibinfo {author} {\bibfnamefont {S. N.}\ \bibnamefont
  {Liu}}, \bibinfo {author} {\bibfnamefont {G. Q.}\ \bibnamefont {Zhang}},
  \bibinfo {author} {\bibfnamefont {L. Z.}\ \bibnamefont {Tang}},\ and\
  \bibinfo {author} {\bibfnamefont {D. W.}\ \bibnamefont {Zhang}},\ }\bibfield
  {title} {\bibinfo {title} {Topological Anderson insulators induced by random
  binary disorders},\ }\href {https://doi.org/10.1016/j.physleta.2022.128004}
  {\bibfield  {journal} {\bibinfo  {journal} {Phys. Lett. A}}\textbf {\bibinfo {volume} {431}},\
  \bibinfo {pages} {128004} (\bibinfo {year} {2022})}\BibitemShut {NoStop}%
\end{thebibliography}

%

\end{document}